\documentstyle[12pt]{article}
\begin{document}
\def\om{\omega}
\def\omt{\tilde{\omega}}
\def\ti{\tilde}
\def\o{\Omega}
\def\bchi{\bar\chi^i}
\def\In{{\rm Int}}
\def\ba{\bar a}
\def\w{\wedge}
\def\ep{\epsilon}
\def\k{\kappa}
\def\Tr{{\rm Tr}}
\def\ST{{\rm STr}}
\def\ss{\subset}
\def\bc{{\bf C}}
\def\br{{\bf R}}
\def\de{\delta}
\def\tr{\triangleleft}
\def\al{\alpha}
\def\la{\langle}
\def\ra{\rangle}
\def\G{\Gamma}
\def\th{\theta}
\def\lm{\lambda}
\def\jp{{1\over 2}}
\def\js{{1\over 4}}
\def\d{\partial}
\def\dz{\partial_z}
\def\dbz{\partial_{\bar z}}

\def\be{\begin{equation}}
\def\ee{\end{equation}}
\def\bea{\begin{eqnarray}}
\def\eea{\end{eqnarray}}
\def\A{{\cal A}}
\def\B{{\cal B}}
\def\g{{\cal G}}
\def\H{{\cal H}}
\def\agh{(\A,\g,\H)}
\def\bT{\bar{\cal T}}
\def\Z{{\cal Z}}
\def\si{\sigma}
\def\*{\ddagger}
\def\j{\dagger}
\def\bz{\bar{z}}
\def\e{\varepsilon}
\def\b{\beta}
\def\ot{\otimes}
\def\bb{\bar b}
\def\De{ (\epsilon V)}
\def\bk{\bar k}
\def\deg{\de^{\g}}
\def\sg{*_{\g}}
\def\vko{V_{\H^{\perp}}}
\begin{titlepage}
\begin{flushright}
{}~
IML 99-02\\
hep-th/9903112
\end{flushright}

\vspace{3cm}
\begin{center}
{\Large \bf A nonperturbative regularization of the supersymmetric
 Schwinger model}\\
[50pt]{\small
{\bf C. Klim\v{c}\'{\i}k}\\ Institute de math\'ematiques de Luminy,
 \\163, Avenue de Luminy, 13288 Marseille, France}

\vspace{1cm}
\begin{abstract}
It is shown that noncommutative geometry is a nonperturbative regulator
 which can manifestly
preserve a space supersymmetry and a supergauge symmetry while keeping
 only finite
number of degrees of freedom in the theory. The simplest $N=1$ case of
the $U(1)$ supergauge
theory on the sphere is worked out in detail.
\end{abstract}
\end{center}
\end{titlepage}
\newpage

\section{Introduction}
 There is a  widespread belief
that "it is unlikely having any regulated form of the theory with the
 supersymmetry
besides string theory" \cite{Dou}.  Nevertheless, we wish to show here,
 though only in the simplest possible
case of the supersymmetric Schwinger model in two dimensions, that there
 exists a regulator
reconciling the supersymmetry with the short distance cut-off. Moreover,
this regulator
has two advantages with respect to strings. First of all, it is genuinely
 nonperturbative
and, secondly, it is  more economic  because it requires only finite number
 of degrees of
 freedom in game.

The mathematical structure which is behind this regulator is noncommutative
 geometry. It is a
 discipline largely developed by A. Connes \cite{Con} and besides
applications in the physics
of the standard model it is expected to have an important impact on a
general structure of quantum
field theory \cite{CoK}. It was actually from the nocommutative rather
than from the supersymmetric side
where the motivation for this work came from. One could witness in the
last few years  a lot
of activity \cite{GKP1,GKP2,Mad,GM,K,Wat,Bal,GP1} concerning model building
 on the so-called
fuzzy sphere. The latter concept was probably invented by Berezin
\cite{Ber}, but the idea to
use it for regularization of  scalar field theories was independently
advocated firstly in \cite{Hop} and \cite{Mad1}.

We should perhaps mention, in order to avoid a misunderstanding, that all
quoted references do not
study the fuzzy sphere as such but rather field theories on it. In the same
 spirit, we may say
that people working on quantum inverse scattering method in two dimensions
have not been studying the two dimensional
Minkowski space but the immensely rich world of two-dimensional field
 theoretical models living on it.
So a question why to devote so much time for studying the fuzzy sphere is
 not well posed; one'd
better ask: "Which theories one can formulate on it?"

There is an encouraging experience  so far that  all interesting field
theoretical structures
can be formulated on the fuzzy sphere. We mean here theories including
fermions \cite{GKP1,Wat,GM,GP1,Bal},
gauge fields \cite{K,Wat,GM,GP1}, topologically nontrivial field
 configurations
\cite{GKP2,Bal} and even those incorporating superscalar fields like
supersymmetric nonlinear $\si$-models \cite{GKP1}. A next step to
 undertake is to formulate
 supersymmetric gauge theories. We shall do it in this paper.

The
 construction of supersymmetric gauge theories is distinguished in
 comparison with the previously studied
models in several aspects. First of all, there is  more structure to
 reconcile with the fuzziness
of the space; besides the supersymmetry, there is the supergauge symmetry
and a need to build up
a complex of superdifferential forms on the noncommutative sphere.
 Secondly, the structure appears very
rigid; while in  the nonsupersymmetric context one usually might advocate
 several nonequivalent
 constructions, in  the supersymmetric context there seems to be a little
room
for ambiguities. Thirdly, but not less importantly, it is quite a
technical task to study the supergauge
theories. A field content of these theories is rich and much work is
 needed  for disentangling
good theories from pathological ones by imposing suitable constraints on
the superfields.

 As it is well-known \cite{WB}, the language of differential forms is
the most natural
one for building up the supersymmetric gauge theories. However, the field
 strength obtained
by applying the coboundary of the supersymmetric complex on the gauge
field (a $1$-form) cannot
be arbitrary but must be, in general, constrained in a way compatible with
 supersymmetry.
Without those constraints, the incovenience  would  be   much stronger
than just an  abundance of additional propagating fields
in the theory. One risks rather the violation of the spin-statistics
theorem, presence of terms containing
four bosonic derivatives and similar serious pathologies.

Unfortunately, the  understanding that the constraints are indispensable
is not sufficient for finding
them. Hitherto there was not found a method which would lead to an
 algorithmic
 identification of the constraints
yielding a good supersymmetric theory. One therefore has to  combine
 an intuition and a lot of calculation to work out the content of the
 theory for a given
set of guessed constraints. As we have alluded already above , in our
 case additional
complications enter the story. Indeed, we need to work always with global
  description of all
involved structures in order to ensure a possibility of a noncommutative
generalization
(non-commutative geometry does not know local charts). Thus,
we have to find a suitable globally defined differential
 complex and a set of constraints which would give a good
theory.

 The guiding principle for seeking the complex which would underlie
 supersymmetric
gauge theories on the  sphere seems clear. One should covariantize the
superderivatives
present in an matter action possessing a global  symmetry to be gauged. If
 we take the standard superscalar
matter \cite{GKP1},
those superderivatives turn out to be a generators of certain superalgebra
 which is
{\it not}
$osp(2,1)$ as one may have expected. What happens is that,
though the resulting supersymmetric theory does turn out to  have the
 $osp(2,1)$ supersymmetry ,
one needs an $osp(2,2)$ covariant structure to uncover it. The reason of
this can be seen already in the
standard case of the flat space (super-Poincar\'e) supersymmetry where one
has to add to the generators
of supersymmetry ($Q,\bar Q$) algebra also the so-called supersymmetric
 covariant derivatives
($D,\bar D$). In the flat
case the derivatives $(D,\bar D)$ are detached from the  superalgebra
 (they anticommute with
the $Q$-generators) and it is of little usefulness to remark that the $Q$'s
 and $D$'s  form
together a $N=2$ superalgebra. However, when  we move from the flat space
 to the sphere, it can be
 seen  (\cite{GKP1})
that the supersymmetric covariant derivatives added to the superalgebra
$osp(2,1)$ entail even
an introduction
of another bosonic generator which completes the structure to that of the
$osp(2,2)$ superalgebra.
 Thus the supersymmetric covariant derivatives  are not detached
from the supersymmetry algebra  and it is natural and, in fact, inevitable
to consider the bigger structure
 which involve them.

In what follows, we shall first introduce an algebra of superfunctions on
the supersphere  following
\cite{K}
and then a differential complex which will underlie the notion of the
supergauge
field. Since the theory of representations of $osp(2,2)$ is not so
notoriously known as that of
$su(2)$ we shall review its basic elements.  In section 3, we first
describe the construction of the supersymmetric Schwinger model
on the ordinary sphere in a way which is most probably also original. Then
we shall give
an invariant description of the complex of the differential superforms.
This invariant description wil
not only render quite transparent the logic of the construction but it will
 prove technically efficient
in formulating the supersymmetric Schwinger model
on the noncommutative sphere. In fact,  formulae  which would appear in the
 noninvariant formulation
in the noncommutative case would be exceedingly cumbersome. We shal finish
with a brief outlook.

\section{A differential complex on the supersphere}
\subsection{Superfunctions on the supersphere}

Consider the algebra of functions on the complex $C^{2,1}$ superplane, i.e.
 algebra
generated by bosonic variables $\bar\chi^{\al},\chi^{\al}, \al=1,2$ and by
fermionic ones $\bar a,a$.
The algebra is equipped with the graded involution \be (\chi^{\al})^{\*}=
\bar\chi^{\al},\quad, (\bar\chi^{\al})^{\*}=\chi^{\al},
\quad, a^{\*}=\bar a,
\quad, \bar a^{\*}=-a,\ee
satisfying  the following properties
\be (AB)^\*=(-1)^{AB}B^\* A^\*,\quad  (A^\*)^\*=(-1)^A A,\ee
and with the super-Poisson bracket
\be \{f,g\}=
\d_{\chi^{\al}}f\d_{\bar\chi^{\al}}g-\d_{\bar\chi^{\al}}f\d_{\chi^{\al}}g +
(-1)^{f+1}[\d_a f\d_{\bar a}g+\d_{\bar a} f\d_a g].\ee
 We can now apply the (super)symplectic reduction with respect
to a moment map $\bar\chi^{\al}\chi^{\al} +\bar a a -1$. The result is a
 smaller algebra
$\A_{\infty}$, that by definition consists of all functions $f$ with the
property
\be \{f,\bar\chi^i\chi^i+\bar a a-1\}=0.\ee
Moreover, two functions obeying (4)
are considered to be equivalent if they differ just by a product
of $(\bar\chi^{\al}\chi^{\al}+\bar a a-1)$ with some other such function.
The smaller algebra $\A_{\infty}$ (the reason for using of the subscript
$\infty$ will become clear soon)
 is referred to as the algebra of superfunctions on the supersphere
\cite{K}. It is sometimes more convenient to work with a different
 parametrization  of $\A_{\infty}$,
using rather the following coordinates
\be z={\chi^1\over\chi^2},
\quad \bar z ={\bar\chi^1\over\bar\chi^2},
\quad b={a\over \chi^2},\quad \bar b={\bar a\over \bar \chi^2}.\ee
 The Poisson bracket (3) then becomes
$$ \{f,g\}=(1+\bz z)(1+\bz z + \bb b)(\dz f \dbz g - \dbz f \dz g)$$
$$ + (1+\bz z)\bb z
((-1)^f\dz f \d_{\bb} g - \d_{\bb} f \dz g) +(1+\bz z) b\bz
(\d_b f \dbz g -
(-1)^f\dbz f \d_b g)$$
\be +(-1)^{(f+1)}(1+\bz z -\bb b \bz z) (\d_b f \d_{\bb}g +
\d_{\bb}f \d_b g). \ee

A natural Berezin integral on $\A_{\infty}$ can be written as
\be I[f]=-{1\over 4\pi^2}
\int d\bar\chi^1\w d\chi^1\w d\bar\chi^2\w d\chi^2 \w
d\bar a \w da ~\delta(\bar\chi^i \chi^i +\bar a a-1) f.\ee
It can be rewritten as
\be I[f]\equiv
-{i\over 2\pi}\int {d\bz \w dz \w d\bb \w db\over 1+\bz z +\bb b} f.\ee
(Note $I[1]=1$.)

Now we are ready to quantize the infinitely dimensional algebra
$\A_{\infty}$ with the goal of obtaining
its (noncommutative) finite dimensional deformation.
The quantization was actually performed in \cite{GKP1}
using the representation theory of $osp(2,2)$ superalgebra.
Here we adopt a different procedure, namely the quantum symplectic
 reduction
 (or, in other words, quantization with constraints).
We start with the well-known quantization of the complex plane $C^{2,1}$.
The generators $\bar\chi^{\al},\chi^{\al},\ba, a$ become creation and
annihilation operators
on the Fock space whose commutation relations are given by the standard
replacement
\be \{.,.\}\to {1\over h}[.,.].\ee
Here $h$ is a real parameter (we have absorbed the imaginary unit into the
 definition
of the Poisson bracket) referred to as the "Planck constant". Explicitely
\be [\chi^{\al},\bar\chi^{\b}]_-=h\delta^{\al\b}, \quad [a,\bar a]_+=h\ee
and all remaining graded
commutators vanish. The Fock space is built up as usual, applying the
creation operators
$\bar\chi^{\al},\ba$ on the vacuum $\vert 0 \rangle$, which is in turn
annihilated
 by the annihilation operators $\chi^{\al},a$.
 We use here the same symbols for the classical and quantum quantities with
 a hope that it will be always
clear from the context which usage we have in mind.

Now we perform the quantum symplectic reduction
with the  moment map
$ (\bar\chi^{\al}\chi^{\al} +\ba a)$. First we restrict the Hilbert space
 only to the vectors
$\psi$ satisfying the constraint
\be (\bar\chi^{\al}\chi^{\al} +\ba a -1)\psi=0.\ee
Hence operators $\hat f$ acting on this restricted space which fulfil
 \be [\hat f, (\bar\chi^{\al}\chi^{\al} +\ba a -1)]=0\ee
 form our deformed version of $\A_{\infty}$.

The spectrum of the operator $(\bar\chi^{\al}\chi^{\al} +\ba a -1)$ in the
 Fock space is
 given by a sequence $Nh-1$, where $N$'s are integers. In order to fulfil
 (11)
for a non-vanishing $\psi$, we observe that the inverse Planck constant
$1/h$ must be an integer $N$. The constraint (xy) then selects only
$\psi$'s living
in the eigenspace $H_N$
of the operator $(\bar\chi^{\al}\chi^{\al} +\ba a -1)$ with the eigenvalue
$0$.
This subspace of the Fock space has the dimension $2N+1$ and the
algebra $\A_N$ of operators(i.e. supermatrices) $\hat f$ acting on it is
 $(2N+1)^2$-dimensional.

When
$N\to\infty$ (the dimension $(2N+1)^2$ then also diverges) we have
the Planck constant approaching $0$ and the algebras $\A_N$ tend
to the classical limit $\A_{\infty}$ \cite{GKP1}.

The Hilbert space $H_N$ is naturally graded. The even subspace $H_{eN}$
is created
 from the Fock vacuum by applying only the bosonic creation operators:
\be (\bar\chi^1)^{n_1}(\bar\chi^2)^{n_2}\vert 0\rangle,\quad n_1+n_2=N,\ee
 while
the odd one $H_{oN}$ by applying both bosonic and fermionic creation
 operators:
\be (\bar\chi^1)^{n_1}(\bar\chi^2)^{n_2} \ba\vert 0\rangle,
 \quad n_1+n_2
=N-1.\ee

 At the level of supercomplex plane $C^{2,1}$, it is the textbook fact
from quantum mechanics that the integral
 $\int d\bar\chi^{\al} d\chi^{\al} d\ba da$ (this is
 the Liouville integral over the superphase space) is replaced under the
quantization
procedure by the supertrace in the Fock space. (The supertrace is the trace
 over
the indices of the zero-fermion states minus the trace over the one-fermion
states).
 The $\delta$ function of the
operator $(\bar\chi^{\al}\chi^{\al} +\ba a -1)$ just restrict the
supertrace to the trace
over the indices of
$H_{eN}$ minus the trace over the indices of $H_{oN}$. Hence an
integration
in $\A_N$ is given by the formula
\be I[\hat f]
\equiv \ST [\hat f],\quad \hat f\in\A_N.\ee
The graded involution $\*$ in the noncommutative algebra $\A_N$ is defined
 exactly as in (1).

\subsection{$osp(2,1)$ and $osp(2,2)$ superalgebras and their
 representations}

The $osp(2,2)$ superalgebra has a convenient basis of even generators
$R_{\pm}$,$R_3$,$\G$ and the odd ones
$V_{\pm},D_{\pm}$, satisfying the following (anti)commutation relations
(see also \cite{SNR,GKP1}):

\be  [R_3,R_{\pm}]=\pm R_{\pm},
\quad [R_+,R_-]=2R_3, \quad [R_i,\G]=0;\ee
\be [D_{\pm},V_{\pm}]_+=0,\quad [D_{\pm},V_{\mp}]_+=\pm{1\over 4}\G;\ee
\be [D_{\pm},D_{\pm}]_+=\mp{1\over 2}R_{\pm},
\quad [D_{\pm},D_{\mp}]_+={1\over 2}R_3;\ee
\be [V_{\pm},V_{\pm}]_+=\pm{1\over 2}R_{\pm},
\quad[V_{\pm},V_{\mp}]_+=-{1\over 2}R_3;\ee
\be [R_3,V_{\pm}]=\pm{1\over 2}V_{\pm},\quad [R_{\pm},V_{\pm}]=0,
 \quad [R_{\pm},V_{\mp}]=V_{\pm};\ee
\be [R_3,D_{\pm}]=\pm{1\over 2}D_{\pm},\quad [R_{\pm},D_{\pm}]=0,
\quad [R_{\pm},D_{\mp}]=D_{\pm};\ee
\be [\G,V_{\pm}]=D_{\pm},\quad [\G,D_{\pm}]=V_{\pm}.\ee
Here and it what follows the commutators are denoted as $[.,.]$ but the
anticommutators have a subscript
$[.,.]_+$. We reserve the notation $\{.,.\}$  for Poisson brackets and
their noncommutative generalizations
(see section 3.2.). If we take a Poisson bracket of two odd elements of
 $\A_{\infty}$ we write
$\{.,.\}_+$.

The superalgebra $osp(2,1)$ is a subsuperalgebra of $osp(2,2)$ generated by
 $R_i,V_{\pm}$.
The irreducible representations of $osp(2,1)$ are classified by one
 parameter, which may be
a positive  integer or a positive  half-integer $j$ and is referred to as
the $superspin$ \cite{SNR}.
Every irreducible $j$-representation is of course a (reducible)
 representation of the $su(2)$-subalgebra
of $osp(2,1)$. Its decomposition into the irreducible components from the
 $su(2)$ point of view
is given by
\be j_{osp(2,1)}=j_{su(2)}\oplus (j-{1\over 2})_{su(2)},\ee
where $j_{su(2)}$ means obviously the standard $su(2)$ spin. The only
exception from the
rule (23) is a trivial superspin zero representation.

The classification of irreducible representations of $osp(2,2)$ is more
involved \cite{SNR}.
There exist two types of them: the  typical and the non-typical ones. The
former are characterized
by the property that they are reducible from the point of view of the
$osp(2,1)$ superalgebra,  while
the latter are irreducible. The typical representation is characterized by
 one positive integer
or half-integer
$j_{osp(2,2)}\geq 1$  called the $osp(2,2)$ superspin
 and by an arbitrary complex number $\gamma\neq \pm 2j$ which is related to
 $\G$ and may be called
 a $\G$-spin. The typical representations considered in this paper will
 have always the $\Gamma$-spin
equal to zero.
They are $8j_{osp(2,2)}$ dimensional
and they have  the following $osp(2,1)$ content
\be j_{osp(2,2)}=j_{osp(2,1)}\oplus (j-1/2)_{osp(2,1)},\ee
hence  the following $su(2)$ content
\be j_{osp(2,2)}=j_{su(2)}\oplus (j-1/2)_{su(2)}\oplus
(j-1/2)_{su(2)}\oplus (j-1)_{su(2)}.\ee

The Lie superalgebra $osp(2,2)$ is naturally represented on the (graded)
commutative associative
superalgebra
$\A_{\infty}$ and also on its noncommutative deformations $\A_N$. In the
 commutative case,
this representation
can be called Hamiltonian since it is generated via the super-Poisson
 bracket (3) by the following
charges
\be r_+=\bar\chi^1\chi^2,\quad r_-=\bar\chi^2\chi^1,\quad r_3=
{1\over 2}(\bar\chi^1\chi^1-\bar\chi^2\chi^2)
\quad \gamma=\bar a a+1\ee
\be 2v_+=\bar\chi^1 a+\bar a\chi^2,\quad 2v_-=\bar\chi^2 a-\bar a\chi^1,
\quad
2d_+=\bar a\chi^2 -\bar\chi^1 a,\quad 2d_-=-\bar\chi^2 a-\bar a\chi^1.\ee
This means that, for instance, $V_+$ acts on an (even) element
$f\in\A_{\infty}$ as
\be V_+f=\{v_+,f\}\ee
and so on for every generator of $osp(2,2)$.

In the non-commutative case, the representation of $osp(2,2)$ is defined by
 the same charges
(26) and (27) but now thought as the operators acting on $H_N$ via scaled
(graded) commutators;
for instance, for even $f$,
\be V_+f=N[v_+,f], \quad f\in\A_N.\ee
The explicit form of the supermatrices $r_i,v_{\al},d_{\al}$ and $\gamma$
 was given in \cite{GKP1}
(eqs. (83)-(91)).

Both the commutative algebra $\A_{\infty}$ and its noncommutative
deformations $\A_N$ are completely
reducible with respect of the $osp(2,2)$ action described above. Their
decompositions
into irreducible components involve only the typical representations and
they are given
explicitely as follows \cite{GKP1}
\be \A_N=\oplus_{j=0}^N j, \quad \A_{\infty}=\oplus_{j=0}^{\infty} j,\ee
where $j$ stands for the $osp(2,2)$ superspin and $j=0$ means the trivial
representation.
The representation space of the latter consists of the constant elements of
 $\A$ and
of the constant multiples of the unit supermatrix in the case of $\A_N$.

The description of the typical multiplet in $\A_{\infty}$ or in $\A_N$ with
 the $osp(2,2)$
 superspin equal to $1$
is easy. In both
cases, the commutative and the noncommutative one, the representation space
 of this superspin $1$
 representation
is spanned by the charges $r_i,\gamma,v_{\pm},d_{\pm}$ which are considered
 as the
elements of $\A_{\infty}$ and $\A_N$, respectively.
Evidently, the $osp(2,2)$-superspin $1$ representation
coincides with the adjoint representation and its dimension is 8. This fact
 has an interesting
consequence, namely that the $\A_N$ valued charges provide also the
$osp(2,2)$
representation with the representation space being $H_N$. This
representation
can be showed to be the $non-typical$ irreducible representation of
$osp(2,2)$
and, as such, it is also the irreducible representation of $osp(2,1)$.
Its $osp(2,1)$ superspin is given by $N/2$.

\subsection{Seeking the  complex}

We could shortly define the differential complex on the supersphere and
then construct
the supergauge theories based on it, but before
doing that it is perhaps desirable to indicate the way how this complex
 was invented. Without those
indications, the interested reader  could  check that the construction
gives correct
results but possibly he would not be convinced that it is somehow unique.

Suppose therefore that we want to construct supergauge theories with the
 underlying superalgebra
being $osp(2,1)$. A natural way to do it consists in trying to
"covariantize" the derivatives
which appear in the action of charged scalar superfield (we are in two
 dimension).
This action  was constructed in \cite{GKP1} and is explicitly given by
\be  S=I[D_+\Phi^\* D_-\Phi-D_-\Phi^\* D_+\Phi+(1/4)\G\Phi^\*\G\Phi],\ee
where $\Phi\in\A_{\infty}$ is a complex scalar superfield on the sphere,
 $I$ is the integral
over $\A_{\infty}$ defined in (7) and the derivatives  $D_{\pm},\G$ were
 defined via the
Poisson bracket (see (29)). If we add those derivatives to  the $osp(2,1)$
 superalgebra
(which acts by means of $R_i,V_{\pm}$) we obtain the $osp(2,2)$
 superalgebra. As was explained
in \cite{GKP1}, the using of sole $osp(2,1)$ generators
was insufficient for constructing a theory respecting the spin-statistics
theorem.

It   seems that our supergauge field multiplet is composed of three
 superfields
$A_{\pm},W$ which one has to add  respectively to three derivatives
$D_{\pm},\G$
in order to covariantize them. The action (31) would then become
$$ S=I[(D_+-A_+)\Phi^\* (D_-+A_-)\Phi-(D_--A_-)\Phi^\* (D_++A_+)\Phi$$
\be +
(1/4)(\G-W)\Phi^\*(\G+W)\Phi].\ee
However, we shall encounter a big trouble in trying
to find a $osp(2,1)$ invariant
field strength corresponding to gauge multiplet $A_{\pm},W$. It seems that
 this field
strenght should be given by an expression that contains only the first
 derivatives
$D_{\pm},\G$ of the multiplet $A_{\pm},W$, for example something like:
\be F=D_+A_- - D_-A_+ +(1/4)\G W.\ee
$F$ defined in this way seems to be nice, since it is indeed $osp(2,1)$
 invariant
(for consistency, the multiplet $A_{\pm},W$ has to transform under the
 $osp(2,1)$ action
in the same way as the derivatives $D_{\pm},\G$ which are transformed
according to (16),(17),
(21) and (22)).
However, the strength so defined is $not$ gauge invariant if we impose the
 evident gauge
 transformation rule
\be A_{\pm}\to A_{\pm}+iD_{\pm}\Lambda, \quad W\to W+i\G\Lambda,\ee
where $\Lambda$ is a real scalar superfield.

Let's continue our search and suppose that the needed field strength is not
 a $osp(2,1)$ singlet
but it is some multiplet with a higher $osp(2,1)$ superspin. However, it is
 not difficult
to show, that no such multiplet exist which would be linear in the
 derivatives  $D_{\pm},\G$
and would respect the gauge transformation (34). The same no go theorem can
 be proved if
we add into the game the derivatives $R_i,V_{\pm}$ acting on $A_{\pm},W$.

Let us therefore add to the
 superspin $1/2$ multiplet $A_{\pm},W$  a superspin $1$ multiplet
of superfields $C_i,B_{\pm}$ whose gauge transformations are defined as
follows
\be C_i\to C_i+iR_i\Lambda, \quad B_{\pm}\to B_{\pm}+iV_{\pm}\Lambda.\ee
Now it  turns out that it exists an  $osp(2,1)$ covariant multiplet of
gauge invariant
field strengths that is linear in the derivatives
 $R_i,V_{\pm},D_{\pm},\G$
and in the superfields $C_i,B_{\pm},A_{\pm},W$. However, the trouble
 reappears:
firstly, it seems to be unnatural to have an abundance of new gauge
superfields in game which even do not
interact with the matter field and which enter only the  pure gauge field
 sector of the Lagrangian.
  Secondly,
and even more importantly, even if we accept that abundance of fields ,
 any polynomial
$osp(2,1)$ invariant Lagrangian built up of that field strenght leads to a
 pathological theory
(higher derivatives, violation of spin-statistics etc.).

It turns out, however, that even this difficulty can be circumvented by
imposing suitable
constraints on the supergauge field multiplet  $A_{\pm},W,C_i,B_{\pm}$
 which would eliminate
the unwanted superfields. This strategy is, of course, standard in the
 superworld but not necessarilly
easy. The subtlety consists in ensuring that  differential constraints in
 the superspace
do not generate differential constraints in the bosonic
variables on the fields which remain in the Lagrangian. At the same time
 one has to ensure that
the constraints are compatible with the $osp(2,1)$ supersymmetry and the
supergauge transformations
(34) and (35). All these conditions are quite stringent and the fact that a
 solution exists even
in the noncommutative case indicates the naturaleness of the compatibility
 of noncommutative geometry
and supersymmetry.

For seeking the good constraints, we adopt a natural assumption that the
constraints are linear
both in the derivatives $R_i,V_{\pm},D_{\pm},\G$
and in the superfields $C_i,B_{\pm},A_{\pm},W$.  The condition of the
compatibility with the
gauge transformations (34)  and (35) selects 32 constraints of that type
which fall into six $osp(2,1)$
supermultiplets. One of these multiplet has the  superspin $1/2$, three of
them  the superspin $1$
and two of them the superspin $3/2$.  The superspin $1/2$ multiplet is
 nothing but the field
strength mentioned above. It reads
\be F_{\pm}=(\G B_{\pm}-V_{\pm}W-2R_{\pm}A_{\mp}+
2D_{\mp}C_{\pm}\mp 2R_3A_{\pm}\pm 2D_{\pm}C_3+2A_{\pm};\ee
\be f=4V_+A_- -4V_-A_+ +4D_-B_+ -4D_+ B_- +2W.\ee

 A tedious (though straightforward) inspection shows that the only viable
constraint is given by  the following superspin $1$ multiplet, i.e.:

\be  \pm 4D_+A_+ +C_{\pm}=0, \quad C_3-2D_-A_+ -2D_+A_-=0;\ee
\be
 B_{\pm}+D_{\pm}W-\G A_{\pm}=0.\ee
 The explicite formulas
for the field strength and the correct constraint will reappear in the
following subsection in terms of the
structures of the differential complex alluded in the introduction. It
 should be clear
that the structure of complex that we  are going to construct is  $implied$
 by our previous
discussion. In other words, we grasp and formalize our search of the
 supersymmetric field strength
and the supersymmetric constraints in terms of that complex.

\subsection{The complex}

We shall describe the differential complex over the supersphere by working
 with the commutative
and noncommutative case at the same time. In fact,
whenever we shall consider the "commutator" in the algebra
 of $\A_N$ we shall have in mind the commutator multiplied by $N$ (for $N$
finite)
and the Poisson bracket (3) for $N=\infty$.

We denote the complex by $\Xi_N$ and we define
it as follows
\be \Xi_N=\oplus_{j=0}^3(\Xi_N)_j.\ee
As usual, the elements of $(\Xi_N)_j$ will be called the $j$-forms. The
spaces  $(\Xi_N)_j$
have the following structure
\be (\Xi_N)_0=(\Xi_N)_3=\A_N;\ee
\be  (\Xi_N)_1=(\Xi_N)_2=\oplus_{i=1}^8 (\A_N)_i,\ee
where $(\A_N)_i=\A_N$ for every value of the index $i$.

Generically, we shall denote by small (capital) Greek characters the
 $0$-forms ($3$-forms)
 and by capital (small) Latin characters the $1$-forms ($2$-forms). Then
 the associative
product $*$ is given by the rules
\be \phi*\psi=\phi\psi, \quad\phi* (A_{\pm},W,C_i,B_{\pm})=
(\phi A_{\pm},\phi W,
\phi C_i,\phi B_{\pm});\ee
\be \phi* (a_{\pm},w,c_i,b_{\pm})=
(\phi a_{\pm},\phi w,\phi c_i,\phi b_{\pm}),
\quad \phi* \Psi=\phi\Psi;\ee
$$ (A^1_{\pm},W^1,C^1_i,B^1_{\pm})* (A^2_{\pm},W^2,C^2_i,B^2_{\pm})=$$
$$(W^1B^2_+ -W^2B^1_+ -2C^1_+A^2_- +2C^2_+A^1_- -2C^1_3A^2_+ +2C^2_3A^1_+,$$
$$W^1B^2_- -W^2B^1_- -2C^1_-A^2_+ +2C^2_-A^1_+ +2C^1_3A^2_- -2C^2_3A^1_-,$$
$$-4B^1_+ A^2_- +4B^1_-A^2_+ -4A^1_- B^2_+ +4A^1_+ B^2_-,$$
$$-4A^1_+ A^2_+,2A^1_- A^2_+ +2A^1_+ A^2_-, 4A^1_-A^2_-,$$
\be W^1A^2_+ -W^2A^1_+, W^1A^2_- -W^2A^1_-);
 \ee
$$ (A_{\pm},W,C_i,B_{\pm})* (a_{\pm},w,c_i,b_{\pm})=$$
\be A_+a_--A_-a_+ +{1\over 4}Ww
-{1\over 2}C_+c_- -{1\over 2}C_-c_+ -C_3 c_3-B_+b_-+B_-b_+.\ee
 The multiplication
of forms by the scalars from the right is defined as in (43) and (44) but
 with $\phi$
standing from the right. The product of a two form with a one-form is given
 as in (46) but
with reversed order of the small and the capital characters. Finally,
all other products are defined to be zero.

It is important to notice, that $A_{\pm},B_{\pm}$ are understood as $odd$
 elements of $\A_N$ while
$C_i,W$ are $even$ (we did not indicate it in (42) in order to avoid too
cumbersome notation).
The same is true for $a_{\pm},b_{\pm}$ and $c_i,w$ respectively.
$\Phi$ and $\psi$ are even .
Of course, these facts play an important role in checking that the product
 (43)-(46) is associative
and, in case of $N=\infty$, also graded commutative.

The coboundary operator $\de$ is given by the
rules
\be \de\Phi=(D_{\pm}\Phi,\G\Phi,R_i\Phi,V_{\pm}\Phi);\ee
$$ \de(A_{\pm},W,C_i,B_{\pm})=$$
$$(\G B_{\pm}-V_{\pm}W-2R_{\pm}A_{\mp}+
2D_{\mp}C_{\pm}\mp 2R_3A_{\pm}\pm 2D_{\pm}C_3+2A_{\pm},$$
$$4V_+A_- -4V_-A_+ +4D_-B_+ -4D_+ B_- +2W,$$
\be -4D_+A_+ -C_+, -C_3+2D_-A_+ +2D_+A_-, +4D_-A_- -C_-,
 -B_{\pm}-D_{\pm}W+\G A_{\pm});\ee
$$\de (a_{\pm},w,c_i,b_{\pm})=$$\be
D_+a_- -D_-a_+ +{1\over 4}\G w-{1\over 2}R_+c_--{1\over 2}R_-c_+
-R_3c_3-V_+b_-+V_-b_+;\ee
\be \de\psi=0.\ee
The operators $R_i,V_{\pm},D_{\pm},\G$ in (47)-(49) act
via the scaled commutators (29) for $N$ finite or via the Poisson
 brackets (3) for $N$ infinite.

One easily checks that the coboundary $\de$ is nilpotent
\be \de^2=0\ee
and
that the $\de$ does verify the
graded Leibniz rule
\be \de (\al*\b)=\de \al *\b +(-1)^{\al}\al *\de \b\ee
in both commutative (infinite $N$) and noncommutative (finite $N$) cases.

 Now we shall give the action
of the  $osp(2,1)$ superalgebra on the complex $\Xi_N$ .
The $osp(2,1)$ action on the $0$-forms is given basically in terms of the
 odd generators $V_{\pm}$.
The action of the bosonic generators $R_i$ can be then derived in terms of
 the anticommutators (19)
of the odd transformations. On the $0$-forms (and $3$-forms), we have the
 following action:
\be \Delta\Phi=(\epsilon_+V_+ +\epsilon_-V_-)\phi
\equiv (\epsilon V)\Phi.\ee
Here $\ep_{\pm}$ are constant Grassmann parameters and $\Delta$ stands for
 an
infinitesimal variation. The parameters $\ep_{\pm}$ behave with respect
 to the graded involution
as follows
\be \ep_+^\*=\ep_-,\quad \ep_-^\*=-\ep_+.\ee
Note that this is the choice of a real form of $osp(2,1)$ with respect
to the graded involution
and it is dictated  by the fact that a $1$-form $(A_{\pm},W,C_i,B_{\pm})$
must fulfil the
following "reality" conditions
\be A_+^\*=A_-,\quad A_-^\*=-A_+,\quad B_+^\*=-B_-,\quad B_-^\*=B_+;\ee
\be C_+^\*=C_-, \quad C_-^\*=C_+,\quad C_3^\*=C_3,\quad W^\*=W,\ee
in order to yield the correct degrees of freedom of a (super)gauge field.

 The action on the $1$-forms (and also on the
$2$-forms) is given as follows
$$\Delta(A_{\pm},W,C_i,B_{\pm})=$$
$$(\De A_+-{1\over 4}\ep_-W,\De A_-+{1\over 4}\ep_+W,
\De W+\ep_+A_+ + \ep_- A_-,$$
$$\De C_+ +\ep_-B_+ ,\De C_- +\ep_+B_-+,
\De C_3 +{1\over 2}\ep_+B_+-\jp\ep_-B_-
,$$
\be \De B_+ -\jp\ep_+C_+ +\jp\ep_-C_3,\De B_-+\jp\ep_+C_3+\jp\ep_-C_-).\ee
It can be easily checked that the coboundary $\de$ is $osp(2,1)$ invariant,
i.e.
\be \Delta\de\om=\de\Delta\om,\quad \om\in\Xi_N.\ee
In fact, Eq.(58) becomes evident when we introduce the invariant
description of the complex in
section 3.2 because the coboundary $\de$ will be given in terms of
invariant operators
of  $osp(2,1)$.
Another important property of $\Delta$ is that it verifies the Leibniz
rule in the complex
$\Xi_N$, in other words
\be \Delta(\om_1*\om_2)=(\Delta\om_1)*\om_2+\om_1*(\Delta\om_2).\ee
This property enables us to construct the $osp(2,1)$ invariants. For
example, the product
of a $1$-form with a $2$-form, given by the formula (46), is  the
$osp(2,1)$ scalar.
Note, however, that the expressions
\be  A_+a_--A_-a_+ +{1\over 4}Ww\ee
and
\be{1\over 2}C_+c_- +{1\over 2}C_-c_+ +C_3^2+B_+b_--B_-b_+,\ee
  are $separately$ $osp(2,1)$ invariant.

The last things we shall need are the "Hodge triangle" $\tr$ and the theory
 of integration on $\Xi_N$.
The Hodge triangle converts an $i$-form in to $(3-i)$-form; it is defined
 simply as the identity
map between $\Xi_0$ and $\Xi_3$, and $\Xi_1$ and $\Xi_2$, respectively.
The integral $I$
will be defined only on the $3$-forms and will be given by (15) for $N$
finite
and by (7) for $N$ infinite.

\section{Field theories}
\subsection{The commutative case}

Let us consider first  the pure gauge theories in the commutative case.
 The gauge field will be a $1$-form $V\in\Xi_N$
subject to the following constraint
\be F\equiv\de V=(.,.,.,0,0,0,0,0),\ee
in words, the first three components of the coboundary $\de V$ are
 unconstrained but the remaining
five have to vanish.
 The reader might have noticed that under the action of the
subalgebra $osp(2,1)$  a generic $1$-form ($2$-form)
decomposes into two multiplets. The first three
components form a multiplet with the $osp(2,1)$-superspin $1/2$ and the
 remaining five with superspin $1$.
 Since  the coboundary $\de$ commutes with the action $\Delta$ of
 $osp(2,1)$,
 it is evident that our constraint (62),
 does respect the $osp(2,1)$ supersymmetry.

The gauge symmetry of the constraint (62) is also obvious due to the
nilpotency of $\de$.
It remains to see whether the constraint involves some unwanted space
derivatives. Fortunately,
it is not the case. Indeed, looking at (48) it is immediately evident that
 the
constraint is resolved with respect to the "additional" superfields
 $C_{\pm},C_3,B_{\pm}$:
\be C_{\pm}=\mp D_{\pm}A_{\pm},\quad C_3=2D_-A_+ +2D_+A_-,\quad
 B_{\pm}=-D_{\pm}W+\G A_{\pm}.\ee
If the field $V$  satisfies the constraint (62)
then the three nonzero components of the field strength $\de V$ will be
 second order expressions
in the derivatives $R_i,V_{\pm},D_{\pm},\G$ acting on the superspin $1/2$
 multiplet
$A_{\pm},W$. This may seem awkward, since a Langrangian
quadratic in the  field strength  will contain terms with four
derivatives. It turns out,
however, that working out the action in $components$ of the superfields
 $A_{\pm},W$
 will  give a non-pathological
action. The same phenomenon takes place in the standard (super-Poincar\'e)
supersymmetric
electrodynamics in two dimensional flat space \cite{Fer} where the kinetic
 term of the
gauge field also contains expressions quartic in the supersymmetric
covariant derivatives
nevertheless the action in components is the standard second-order  one.

Let us write a pure gauge field action on the $commutative$ supersphere as
 follows
\be S_{\infty}(V)=I[\alpha'\de V*\tr \de V+\b' V* \de V],
\quad V\in\Xi_N.\ee
Where $\alpha'$ and $\b'$ are real parameters and  the components of
 $V=(A_{\pm},W,C_i,B_{\pm})$
 are supposed to satisfy the reality
conditions (55) and (56). It is evident that the action is gauge invariant
with respect
to the transformation \be V\to V+i\de\Lambda,\ee
where $\Lambda$ is a $0$-form. Of course, $V$ is to satisfy the constraint
 (62),
hence the components $C_i,B_{\pm}$ are given by (63). Having in mind this,
we can evaluate the coboundary of $V$:
\be \de V=(F_+,F_-,f,0,0,0,0,0),\ee
where
$$F_{\pm}=(\G^2+2)A_{\pm}-(\G D_{\pm} +
V_{\pm})W\mp 12D_{\mp}D_{\pm}A_{\pm}\pm 12D_{\pm}^2A_{\mp};$$

\be f=2W+4(D_+D_--D_-D_+)W+4(V_+-D_+\G)A_- -4(V_--D_-\G)A_+.\ee

Before giving  a noncommutative version of the action (64), let us
write its content in a more familiar parametrization. Set
\be A_+=\jp (A-\bz\bar A)+\jp d_+ K,\quad A_-=-\jp(zA+\bar A)+\jp d_- K,
\quad W=\bb\bar A-bA+\gamma K,\ee
where $d_{\pm},\gamma$ were defined in (26) and (27). The components
$F_{\pm},f$ then become
$$ F_+=-{3\over 2}n[-\bar DDA +\bz \bar DD\bar A+\bb(D\bar A+\bar DA)+
D^2\bar A+\bz\bar D^2 A]$$
\be +2d_+n(\bar DA+D\bar A)-4d_+K+2nd_+\bar DDK-2(D+\bz \bar D)K;\ee
$$ F_-={3\over 2}n[-z\bar DDA -\bar DD\bar A+b(D\bar A+\bar DA)+
zD^2\bar A-\bar D^2 A]$$
\be +2d_-n(\bar DA+D\bar A)-4d_-K+2nd_-\bar DDK-2(\bar D-z D)K;\ee
$$f=3n[\bb(\bar DD\bar A)-b(\bar DDA)-2(D\bar A+\bar DA)+bD^2\bar A+
\bb\bar D^2A]$$
\be 2\gamma n(\bar DA+D\bar A)+2\gamma n\bar DDK+
4(\bb\bar D+bD)K-4\gamma K.\ee
Here $ n=1+\bz z+\bb b$ and the operators $D,\bar D$ are the standard
supersymmetric covariant derivatives in two dimensions, i.e.
\be D=\d_b+b\dz,\quad \bar D=\d_{\bb}+\bb\dbz.\ee
It is perhaps worth giving formulae that express the derivatives
$R_i,\G,D_{\pm},V_{\pm}$ in terms
of the derivatives $D,\bar D$ and $Q,\bar Q$, where
\be  Q=\d_b-b\dz,\quad \bar Q=\d_{\bb}-\bb\dbz.\ee
Here they are:
\be D_+=\jp(D-\bz\bar D),\quad D_-=-\jp(\bar D+zD);\ee
\be  V_+= \jp (Q+\bz\bar Q),\quad V_-=\jp(\bar Q
-zQ);\ee
\be \G=\bb\d_{\bb}-b\d_b,\quad R_3=\bz\dbz-z\dz+
\jp\bb\d_{\bb}-\jp b\d_b;\ee
\be R_+=-\dz -\bz^2\dbz-\bz\bb\d_{\bb},\quad R_-=\dbz +z^2\dz+zb\d_b.\ee

Using the formulae (46),(68) and (69) -(71), we obtain the action (64) in
 terms of $A,\bar A, K$.
The explicit formula is somewhat cumbersome and we do not list it here
since it is pathological!
The trouble is caused by the superfield $K$ that is contained in the action
 in the form $(\bar DDK)^2$.
This gives the bosonic derivatives of fourth order. Fortunately, there is a
 manifestly supersymmetric
and gauge invariant way for getting rid of the unwanted field $K$. Indeed, a
constraint
\be K=0\ee
simply does the job. Having in mind the  noncommutative generalization, it
 is desirable to  formulate
this additional constraint in terms of the fields $A_{\pm},W$. It reads
\be d_+A_--d_-A_++\js\gamma W=0.\ee
It is not difficult to verify the gauge symmetry (78) or (79) and
the $osp(2,1)$ supersymmetry of (79).

After imposing the additional constraint (79), the action (64)  becomes
\be S_{\infty}={1\over 2\pi i}\int d\bz dz d\bb db
\{\alpha \bar D(n\omega)D(n\omega)+
\b n\omega^2\},\ee
where
\be \omega=\bar DA +D\bar A, \quad n=1+\bz z+\bb b\ee
and the parameters $\alpha,\b$ are linear combinations of $\alpha',\b'$.

It is instructive to see how the $osp(2,1)$ superinvariance is realized in
 this parametrization
of the gauge field multiplet. We have
\be \Delta A=(\ep_+V_+ +\ep_-V_-)A+\jp\ep_-bA;\ee
\be \Delta \bar A=(\ep_+V_+ +\ep_-V_-)\bar A-\jp\ep_+\bb\bar A\ee
and therefore
\be \Delta\omega=(\ep_+V_+ +\ep_-V_-)\omega+\jp(\ep_-b-\ep_+\bb)\omega,\ee
or, equivalently,
\be \Delta(n\omega)=(\ep_+V_+ +\ep_-V_-)(n\omega).\ee
From the last formula (85), the $osp(2,1)$ supersymmetry of the action
 (80) immediately follows.
The gauge symmetry $A\to A+iD\Lambda$,$\bar A\to \bar A+i\bar D\Lambda$ is
also evident.

We are now ready to cast the action (80) in components. Set
\be iA=\zeta+bv+\jp\bb{(w+iu)\over 1+\bz z} +
\bb b({\eta\over 1+\bz z}-\dz\zeta);\ee
\be i\bar A=\bar\zeta-\jp b{(w-iu)\over 1+\bz z}+\bb\bar v +
\bb b({\bar\eta\over 1+\bz z}
-\dbz\bar\zeta).\ee
Here $u,w$ are real, $v$ and $\bar v$ mutually complex conjugate and
 $\eta^{\*}=\bar\eta$,
$\zeta^{\*}=\bar\zeta$.
We calculate
\be in(\bar DA+D\bar A)=iu+b\eta-\bb\bar\eta+\bb b
[(1+\bz z)(\dbz v-\dz \bar v)+{iu\over 1+\bz z}].\ee
Thus the fields $\zeta,\bar\zeta$ and $w$ have dropped out.

It turns out that an avoiding a mutual coupling of the fields $u$ and $v$
 in the action requires setting
$\b=-2\alpha$ in the action (80). Thus we respect
this most natural choice and  write finaly
$$S_{\infty}= {-\al\over 2\pi i}
\int d\bz dz \{-(1+\bz z)^2(\dbz v-\dz \bar v)^2+\dbz u\dz u +
{u^2\over (1+\bz z)^2}$$
\be +\eta\dbz\eta+
\bar\eta\dz\bar\eta
+4{\bar\eta\eta\over (1+\bz z)}\}.\ee

  Note that the action (89) {\it differs} from  that of the standard free
 supersymmetric electrodynamics
in the flat Euclidean space \cite{Fer} by the presence of the mass term for
 the dynamical fields $u$ and
$\eta,\bar\eta$. The reader may say that it is not surprising that
theories
on different manifolds have different actions. Note, however, that  the
 matter action (32),
 rewritten
in terms of the superfields $A,\bar A$  and a complex matter superfield
 $\Phi$,
is exactly of the same form as that of the matter sector of supersymmetric
Schwinger model in the flat two-dimensional space \cite{Fer}:
\be S_{matter}={1\over 2\pi i}\int d\bz dz d\bb db[(\bar D-\bar A)\Phi^\*
(D+A)\Phi
+(\bar D+\bar A)\Phi(D-A)\Phi^\*].\ee
 Of course, this coincidence (due to superconformal invariance of the
 massless superscalar matter in two
dimensions) is only
formal because, in spite of the same form of the action, the superfields
 belong to {\it different}
algebras! The matter superfield $\Phi$ is an element of the algebra of the
 superfunctions
on the supersphere while in the flat case it would be an element of the
algebra of superfunctions
on the flat Euclidean superspace.

\subsection{An invariant description of the complex}

The complex constructed in section 2.4 looks somewhat cumbersome and we may
 wonder
whether it exists its description which would be more elegant and natural.
 It turns out, that the
answer to this question is positive; in fact, we shall present the complex
 in a way which does not
need to choose a basis in the superalgebra $osp(2,2)$ or a basis in the
 space of $1$-forms.
The invariant picture is entirely based on standard structures of super-Lie
 algebras and
it is not only more esthetic but also it is efficient for technical
 purposes. Indeed, the noncommutative
generalization of the commutative theories of the previous subsection
 requires adding a quadratic
term in the definition of the field strength and  cubic and quartic terms
in the action functional
of the field theory. Consulting the formula (45) the reader may easily
convince himself that working with
the product of forms in the previous non-invariant description would be
 very messy. However, the invariant
picture will enable us readily to formulate the constraint (62) and even to
 solve it explicitely!
What is even more remarkable in the story that the complex can be
constructed for a much general
class of Lie super-algebras than just $osp(2,2)$.
Here are the details:

\vskip1pc
\noindent {\bf Definition 1}: A super-Poisson algebra $\A=\A_0\oplus \A_1$
is a $Z_2$-graded associative algebra over
the field of complex numbers  $\bc$
   equipped with the structure
 of  a super-Lie algebra with an even (super-Poisson) bracket $\{.,.\}$
compatible with the associative
multiplication $m:\A\otimes\A\to\A$; i.e.
\be \{X,YZ\}=\{X,Y\}Z+(-1)^{XY}Y\{X,Z\}.\ee
Moreover,  it is required that $\A$ possess an even unit element $e$
such that $eX=Xe=X$ and
$\{e,X\}=0$  for all $X\in \A$ .  Finally, $\A$ is equipped with a linear
 superstrace
 $\ST :\A \to \bc$, in particular $\ST(e)=1$.
 The supertrace is supposed to vanish
for  Poisson  brackets: $\ST\{X,Y\}=0; X,Y\in \A$  and also for odd
elements: $\ST(\A_1)=0$.

\vskip1pc
Let us take as an example of $\A$ the algebra $\A_{\infty}$
of the superfunctions on the supersphere with the super-Poisson bracket
given by (3) or (6) and
the supertrace given by the integral (7). Another example is the
 noncommutative algebra
of $(2N+1)\ot(2N+1)$-supermatrices , denoted as $\A_N$ in section 2.1. This
 algebra defines the
fuzzy supersphere \cite{GKP1} and it is the noncommutative deformation
(or Berezin quantization)
of the algebra $\A_{\infty}$ with the value of the "Planck constant"
 $h=1/N$.
 It should be therefore no surprise that the Lie bracket in $\A_N$ is not
just the ordinary commutator
inherited from the associative multiplication in $\A_N$ but the commutator
 multiplied by $N$
(playing role of the inverse Planck constant):
\be \{X,Y\}=Nm(X\ot Y)-Nm(Y\ot X)=N[X,Y],\quad  X,Y\in \A_N.\ee
The definition of the bracket with this normalization is crucial for
verifying normalization of all
formulae in this paper, in particular the important constrain (113).
Note that  we denote the super-Lie bracket in $\A_N$ by the same symbol as
 in $\A_{\infty}$. The
reason is that the former gives the latter in the commutative limit
 $N\to\infty$. Finally, the
 supertrace $\ST$
in $\A_N$ is nothing but the standard supertrace over
 $(2N+1)\ot(2N+1)$-supermatrices.
It is evident from (13) and (14), that $\ST$ is correctly normalized, which
 means that the $\ST$ of the
unit supermatrix  is equal to $1$.
\vskip1pc

\noindent {\bf Definition 2}: We say that $(\A,\g)$ is a supersymmetric
 double over a super-Poisson
algebra $\A$, if $\g=\g_0\oplus \g_1$ is a super-Lie subalgebra of $\A$
(but not necessarilly
the associative
subalgebra of $\A$!)
and a bilinear form $\ST \circ m$
restricted to $\g$
is non-degenerate. In this case the bilinear form $\ST\circ m$ determines
 an
 element $C^{\g}\in \g\ot \g$
called a quadratic Casimir element of the double $(\A,\g)$.

\vskip1pc
Now we construct a canonical complex $\o(\A,\g)$ over the double $(\A,\g)$
 as follows
\be \o(\A,\g)=\oplus_{i=0}^3\o_i(\A,\g),\ee
where
$$ \o_0(\A,\g)=\o_3(\A,\g)=(\A_P)_0\equiv e\ot (\A_P)_0,$$
 \be \o_1(\A,\g)=\o_2(\A,\g)=
(\g_0\ot (\A_P)_0)\oplus (\g_1\ot (\A_P)_1).\ee
Here the notation $\A_P$ means that one does not consider the algebra $\A$
 over the field $\bc$
but one replaces $\bc$ by a graded commutative algebra $P$. The subscript
 $0$ in
$(\A_P)_0$ ($1$ in
$(\A_P)_1$) then means that  one considers only a subspace of $\A_P$ which
 consists of all elements
of $\A_P$ even (odd) with respect to the sum of gradings of $\A$
 and of $P$.

\vskip1pc

\noindent $Note$: Considering the algebra $\A$ over $P$ instead over the
field of complex numbers
$\bc$ is not an  useless complication,
but it is dictated by the field theoretical applications. For instance,
any element $f(z,\bz,b,\bb)\in\A_{\infty}$ can be expanded in the Taylor
series in $b,\bb$. The term proportional
to $b$ is of the form $\psi(z,\bz)b$. But $\psi(z,\bz)$, being a fermion,
 is not a $\bc$-valued function
! It should be Grassmann-valued  which means that it should be valued in
 odd part $P_1$ of some
$Z_2$-graded commutative algebra $P$. All  this is the standard
supersymmetric  story
so we do not give more  details here.

\vskip1pc
In order $\o(\A,\g)$ be a complex we need to introduce a coboundary
operator $\de^{\g}:
 \o_i(\A,\g)\to\o_{i+1}(\A,\g)$ such that $(\deg)^2=0$ and an
associative product $*_{\g}: \o_i(\A,\g)\ot\o_j(\A,\g)\to\o_{i+j}(\A,\g)$
 compatible with $\de^{\g}$.
By the compatibility we mean, of course, the Leibniz rule

\be \de^{\g}(X^i\sg Y^j)=\deg X^i\sg Y^j + (-1)^i X^i\sg \deg Y^j,
\quad X^i\in \o_i(\A,\g),
Y^j\in \o_j(\A,\g) .\ee

In order to give a simple description of $\deg$ and $\sg$, we note that
$\A\ot\A$ naturally acts
on itself in one of four ways: $m\ot m$,$m\ot ad$,$ad\ot m$ and $ad\ot ad$;
 e.g.for $X,Y\in \A$
we have
\be X_{m\ot ad}(Y)
\equiv (-1)^{Y^{(1)}X^{(2)}}X^{(1)}Y^{(1)}\ot \{X^{(2)},Y^{(2)}\},\ee
where $X=X^{(1)}\ot X^{(2)}$ and $Y=Y^{(1)}\ot Y^{(2)}$.
Now we define $\deg$ as follows,
\be \deg X^0=C^{\g}_{m\ot ad}X^0;\ee
\be  \deg X^1=C^{\g}_{ad\ot ad}X^1 + \jp d_{\g}X^1;\ee
\be \deg X^2=ad X^2\equiv \{X^{2(1)},X^{2(2)}\};\ee
\be \deg X^3=0,\ee
where $X^i\in \o_i(\A,\g)$ and $d_{\g}$ is a "Dynkin" number which can be
 defined by the relation
\be \ST (XY)= {1\over d_{\g}}{\rm sTr}(adXadY).\ee
Here sTr is the  superstrace over supermatrices in adjoint representation
 of the Lie superalgerba
$\g$.
The multiplication $\sg$ is given by the following table
\be \left(\matrix{X^i\sg Y^j&Y^0&Y^1&Y^2&Y^3\cr
 X^0&m\ot m&m\ot m&m\ot m&m\ot m
\cr
X^1&m\ot m&ad\ot m&(\ST\ot {\rm Id})(m\ot m)&0\cr
x^2&m\ot m&(\ST\ot {\rm Id})(m\ot m)&0&0\cr X^3&m\ot m&0&0&0 }\right).\ee
For example,
\be X^1\sg Y^1=X^1_{ad\ot m}(Y^1),\ee
or
\be X^1\sg Y^2= (-1)^{Y^{(1)}X^{(2)}}X^{1(2)}Y^{2(2)}
\ST(X^{1(1)}Y^{2(1)}).\ee

\vskip1pc
\noindent{\bf Definition 3}: We say that $(\A,\g,\H)$ is a supersymmetric
 triple, if
it exists a subspace  $\H$ of $\A$ such that

\noindent 1) $\H$ is a super-Lie subalgebra of $\g$;

\noindent 2)  $(\A,\H)$ is
the supersymmetric double with the Casimir element $C^{\H}\in\H\ot\H$,
coboundary $\de^{\H}$
and product $*_{\H}$;

\noindent 3) An element $C\equiv C^{\g}-C^{\H}$ fulfils $m(C)\in\bc e$;

\noindent 4) $ad(\H^{\perp}\ot \H^{\perp})\ss \H$, where $\H^{\perp}$ is an
orthogonal complement of $\H$ in $\g$ with respect to $\ST\circ m$.

\vskip1pc

Now we construct a canonical complex $\o(\A,\g,\H)$ over the triple
 $(\A,\g,\H)$ as follows
\be \o(\A,\g,\H)=\oplus_{i=0}^3\o_i(\A,\g,\H),\ee
where
$$ \o_0(\A,\g,\H)=\o_3(\A,\g,\H)=(\A_P)_0\equiv e\ot (\A_P)_0,$$
 \be \o_1(\A,\g,\H)=\o_2(\A,\g,\H)=
(\g_0\ot (\A_P)_0)\oplus (\g_1\ot (\A_P)_1).\ee
We define the exterior derivative $\de$ on $\o\agh$ as follows,
\be \de X^0=\deg X^0, \quad \de X^2=\deg X^2, \quad  \de X^3=\deg X^3;\ee
\be \de X^1=\deg X^1-\de^{\H}X^1_{\H},\ee
where  $X^i\in \o_i(\A,\g,\H)$ and $X^1_{\H}$ means the orthogonal
projection of $X^1$ from $\g\ot \A$ into $\H\ot A$.

A product $*$ in $\o\agh$ is defined by the same table as the product $\sg$
in $\o(\A,\g)$, except
of the multiplication of $1$-forms where we have
\be X^1*Y^1=X^1\sg Y^1-X^1_{\H}*_{\H}Y^1_{\H}.\ee
 It is straightforward exercise to check that the product $*$ and
 the coboundary $\de$ verify the
Leibniz rule.

\subsection{Noncommutative supersymmetric electrodynamics}
Suppose that the super-Poisson algebra $\A$ is such that its Lie bracket
$\{.,.\}$ is derived from its
associative multiplication i.e.,
as an $N$-multiple of its commutator  like in (92). Then
a noncommutative pure supersymmetric electrodynamics  over $\A$ is a theory
 of $1$-forms in the
 triple complex
 $\o\agh$ defined by an action
\be S={1\over g^2}\ST[\alpha'\tr F*F+\b'( V*\de V+{2\over 3}V*V*V)], \ee
where
\be F\equiv \de V+ V*V\ee
is  the field strength  of $V$, $\alpha'$,$\beta'$ two real parameters
(cf. (64)),
$g$ a coupling constant and the Hodge triangle  $\tr $
is the identity map between
 $\o_1\agh$ and  $\o_2\agh$. Moreover, $V$ is considered to be a real
$1$-form
$V^\*=V$  subject to two constraints
\be (\de V+V*V)_{\H}=0\ee
and
\be C*\vko+ V_{\H^{\perp}}*C+{1\over N}\tr \vko *\vko=0.\ee
Here $\vko$ means the orthogonal projection of $V$ to $\H^{\perp}$ and $C$
is viewed as a $2$-form.
The graded star $\*$ on the complex is defined by means of the graded star
 $\*$ on the algebra $\A$.
For example,
\be (X^1)^\*=(X^{1(1)})^\*\ot(X^{1(2)})^\*, \quad X^1\in \o_1\agh.\ee

The action $S$ and both constraints are invariant with respect to

\vskip1pc
\noindent  1) gauge transformations
\be V\to UVU^{-1}-(\de U)U^{-1}, \quad U^{-1}=U^{\*};\ee
\noindent 2) $\H$-supersymmetry\footnote{We mean here the  real form of
 $\H$-superalgebra
which respects the reality of the $1$-form $V$; cf. (54).}, where
$h\in \H$ acts on $\o\agh$ as follows
\be h(X^{0,3})=ad(h\ot X^{0,3});\ee
\be h(X^{1,2})=(e\ot h)_{m\ot ad}X^{1,2}+(h\ot e)_{ad\ot m}X^{1,2}.\ee
The constraint $(\de V+V*V)_{\H}=0$ can be solved explicitely, thanks to
the assumption (4) in the definition
of the supersymmetric triple $\agh$ (the assumption (3) is needed for the
gauge invariance
of the constraint (113)). The solution reads
\be V_{\H}={2\over d_{\H}-d_{\g}}(C_{ad\ot ad}V_{\H^{\perp}}+
V_{\H^{\perp}}*V_{\H^{\perp}}).\ee
Thus we insert  $V_{\H}$ from (118) into (110) and we obtain a theory
 containing
only $1$-forms  $V_{\H^{\perp}}$.

 An interaction with matter can also
be expressed in terms of the complex $\o\agh$. Let $\Phi$ be a complex
 $0$-form, then
\be S_{matter}=\ST[((\de^{\g}-\de^{\H}+V_{\H^{\perp}} )\Phi)^\* *
\tr(\de^{\g}-\de^{\H}+V_{\H^{\perp}})\Phi].\ee
If we add $S_{matter}$ to $S$ in (110)  we obtain the $\H$-supersymmetric
Schwinger model over the
Poisson algebra $\A$.

For the commutative resp. noncommutative fuzzy supersphere
 ($\A=\A_{\infty}$
resp. $\A=\A_N$) and the superalgebras $\H=osp(2,1)$
and $\g=osp(2,2)$, the complex $\o\agh$ is isomorphic to the complex
 $\Xi_{\infty}$ resp.
$\Xi_N$ of  the section 2.4. We present few formulae connecting the two
presentations of the same
complex. First of all the algebras $\g$ and $\H$ are generated by the
Hamiltonians
(26) and (27) in both commutative and noncommutative cases.
In what follows, the parameter $N$ will stand for  either a finite integer
 (the noncommutative
case) or $\infty$ (the commutative one). Thus
\be \vko=+2d_-\ot A_+ -2d_+\ot A_-  -\jp \gamma\ot W
\equiv (A_+,A_-,W,0,0,0,0,0);\ee
\be F=+2d_-\ot F_+ -2d_+\ot F_-  -\jp \gamma\ot f
\equiv (F_+,F_-,f,0,0,0,0,0);\ee
\be C=+2d_-\ot d_+ - 2d_+\ot d_- -\jp\gamma\ot\gamma;\ee
\be d_{\g}={4\over 1+1/N}, \quad d_{\H}={6\over 1+1/N}.\ee
Note also that $C^\*=C$ and the condition $\vko^\*=\vko$ does indeed
reproduce the reality conditions
 (55) and (56).
It is trivial to check that the conditions 3) and 4) of the definition 3
 are indeed fulfilled.
For this, one uses the relations (26)-(27), (122) and (16) -(22).

The main message of this paper is that the
Schwinger model formulated in this section for finite $N$ becomes in the
limit $N\to\infty$
the standard commutative supersymmetric
theory of section 3.1. In particular, the actions (110)  and (119) become
(64) and (90)
respectively and the constraints (112) and (113)
 become
(62) and (79) in this limit. Moreover, the gauge transformation (115)
becomes the gauge transformation
(34) and (35) and the supersymmetry transformations (116) and (117)
 become
the transformations (53) and (57). It is in this sense that we consider the
 theory for finite $N$
as the nonperturbative regularization of the standard commutative theory.

 Note  the presence
of the expressions like $\de V+V*V$ in our theory which are characteristic
 for non-abelian gauge theories.
They appear  because of the noncommutativity of the fuzzy sphere, but in
the commutative limit
the terms like $V*V$ disappear and we are left with the abelian theory. In
 fact,
 the coboundary $\de$ commute
 only with one parametric subgroup
of the gauge group, consisting of the elements of the form
 $U=\exp{(i\alpha)}e$ where $e$ is the unit element
of $\A$.  This is another sign that we deal with the noncommutative
 deformation of an $U(1)$ gauge theory.
The reader might appreciate also the economy of using the invariant
formulation for writing the
action functionals. In fact, the already cumbersome formulae of
 section 3.1, which are written
in the noninvariant language,  would
be even much more cumbersome in the noncommutative case due to the presence
 of the $V*V$ terms.

\section{Conclusions and outlook}
We have constructed the supersymmetric Schwinger model on the
 noncommutative sphere. The theory
possess only finite number of degrees of freedom nevertheless  it is
manifestly supersymmetric and
supergauge invariant.
 The basic structural ingredient of the construction of the model is the
complex $\o(\A,\g,\H)$ for $\g=osp(2,2)$ and  $\H=osp(2,1)$. It is
remarkable that the complex
can be constructed for large class of superalgebras hence we expect that
 supergauge theories
could be in a similar fashion  constructed for higher dimensional
projective spaces. It is  also not
difficult
 to suggest
a generalization to more general gauge groups than $U(1)$, however, due to
the amount of work
necessary for doing it we prefer to postpone it for a later
publication.

Note finally that $1$-forms in $\o(A,\g)$
can be interpreted as $1$-cochains in the Lie superalgebra cohomology but
 $1$-forms  in
$\o\agh$ are $not$ relative cochains modulo $\H$. This suggests that it
 may exists an interesting
variant of the Lie superalgebra cohomology of $\g$ with respect to $\H$.

\end{document}